\documentclass[%
 reprint,
superscriptaddress,
 amsmath,amssymb,
 aps,
]{revtex4-2}

\usepackage{graphicx}% Include figure files
\usepackage{dcolumn}% Align table columns on decimal point
\usepackage{bm}% bold math
\usepackage{bbold}% bold math
\usepackage{hyperref}% add hypertext capabilities
\usepackage{stackengine}
\usepackage[english]{babel}
\begin{document}

\preprint{APS/123-QED}

\title{Variational preparation of entangled states in a system of transmon qubits}% Force line breaks with \\

\author{Alexander Yeremeyev}
\email{Alexander.Yeremeyev@skoltech.ru}
\affiliation{Center for Engineering Physics, Skolkovo Institute of Science and Technology, 121205 Moscow, Russia}
\affiliation{Laboratory of Artificial Quantum Systems, Moscow Center for Advanced Studies, 123592 Moscow, Russia}
\author{Aleksei Tolstobrov}
\affiliation{Center for Engineering Physics, Skolkovo Institute of Science and Technology, 121205 Moscow, Russia}
\affiliation{Laboratory of Artificial Quantum Systems, Moscow Center for Advanced Studies, 123592 Moscow, Russia}
\author{Gleb Fedorov} 
\affiliation{Laboratory of Artificial Quantum Systems, Moscow Center for Advanced Studies, 123592 Moscow, Russia}
\affiliation{Kotel'nikov Institute of Radio Engineering and Electronics, Russian Academy of Sciences, 125009 Moscow, Russia}
\author{Shtefan Sanduleanu} 
\affiliation{Laboratory of Artificial Quantum Systems, Moscow Center for Advanced Studies, 123592 Moscow, Russia}
\affiliation{Kotel'nikov Institute of Radio Engineering and Electronics, Russian Academy of Sciences, 125009 Moscow, Russia}
\author{Peter Shlykov} 
\affiliation{Center for Engineering Physics, Skolkovo Institute of Science and Technology, 121205 Moscow, Russia}
\affiliation{Laboratory of Artificial Quantum Systems, Moscow Center for Advanced Studies, 123592 Moscow, Russia}
\author{Sergey Samarin} 
\affiliation{Center for Engineering Physics, Skolkovo Institute of Science and Technology, 121205 Moscow, Russia}
\affiliation{Laboratory of Artificial Quantum Systems, Moscow Center for Advanced Studies, 123592 Moscow, Russia}
\author{Shamil Kadyrmetov} 
\affiliation{Laboratory of Artificial Quantum Systems, Moscow Center for Advanced Studies, 123592 Moscow, Russia}
\author{Artyom Muraviev} 
\affiliation{Center for Engineering Physics, Skolkovo Institute of Science and Technology, 121205 Moscow, Russia}
\affiliation{Laboratory of Artificial Quantum Systems, Moscow Center for Advanced Studies, 123592 Moscow, Russia}
\author{Aleksey Bolgar} 
\affiliation{Laboratory of Artificial Quantum Systems, Moscow Center for Advanced Studies, 123592 Moscow, Russia}
\author{Daria Kalacheva} 
\affiliation{Laboratory of Artificial Quantum Systems, Moscow Center for Advanced Studies, 123592 Moscow, Russia}
\author{Viktor Lubsanov} 
\affiliation{Laboratory of Artificial Quantum Systems, Moscow Center for Advanced Studies, 123592 Moscow, Russia}
\author{Aleksei Dmitriev} 
\affiliation{Laboratory of Artificial Quantum Systems, Moscow Center for Advanced Studies, 123592 Moscow, Russia}
\affiliation{Kotel'nikov Institute of Radio Engineering and Electronics, Russian Academy of Sciences, 125009 Moscow, Russia}
\author{Evgenia Alekseeva} 
\affiliation{Laboratory of Artificial Quantum Systems, Moscow Center for Advanced Studies, 123592 Moscow, Russia}
\author{Oleg V. Astafiev} 
\affiliation{Center for Engineering Physics, Skolkovo Institute of Science and Technology, 121205 Moscow, Russia}%
\affiliation{Laboratory of Artificial Quantum Systems, Moscow Center for Advanced Studies, 123592 Moscow, Russia}

\date{\today}% It is always \today, today,
             %  but any date may be explicitly specified

\begin{abstract}
The conventional method for generating entangled states in qubit systems relies on applying precise two-qubit entangling gates alongside single-qubit rotations. However, achieving high-fidelity entanglement demands high accuracy in two-qubit operations, requiring complex calibration protocols. In this work, we use a minimally calibrated two-qubit iSwap-like gate, tuned via straightforward parameter optimization (flux pulse amplitude and duration), to prepare Bell states and GHZ states experimentally in systems of two and three transmon qubits. By integrating this gate into a variational quantum algorithm (VQA), we bypass the need for intricate calibration while maintaining high fidelity. Our proposed methodology employs variational quantum algorithms (VQAs) to create the target quantum state through imperfect multiqubit operations. Furthermore, we experimentally demonstrate a violation of the Clauser–Horne–Shimony–Holt (CHSH) inequality for Bell states, confirming their high fidelity of preparation.
\begin{description}
\item[Keywords]
variational quantum algorithms, transmon qubits, Bell states, GHZ state,  CHSH inequality violation.
\end{description}
\end{abstract}

%\keywords{Suggested keywords}%Use showkeys class option if keyword
                              %display desired
\maketitle

%\tableofcontents

\section{\label{sec:level1} Introduction}

Recent research in quantum technologies has seen growing interest in applying machine learning to experiments with various quantum systems. Machine learning has demonstrated potential for qubit control through pulse optimization, enabling tasks such as multi-qubit state preparation \cite{Brown_2023} and single-qubit gate optimization \cite{Genois2024quantumoptimalcontrolsuperconducting}.

For noisy intermediate-scale quantum systems \cite{Preskill2018quantumcomputingin}, where the limited number of qubits and the absence of error correction prevent direct realization of quantum advantage, variational quantum algorithms (VQAs) have been successfully applied to tasks such as state preparation \cite{Sagastizabal_2021}, classification \cite{Rist_2017, Havl_ek_2019, Dutta_2022, Herrmann_2022,  Tolstobrov_2024_Phys_Rev_A, Tolstobrov_2024_RF}, image recognition \cite{Ren_2022, Tolstobrov_2024_Phys_Rev_A},  and simulating quantum systems \cite{Pan_2023}. Recently, variational quantum algorithms have been theoretically explored as a tool for generating multi-qubit entangled states \cite{hai2023variationalpreparationentangledstates}. However,  the scheme discussed in that work does not seem to be practically feasible because it relies on the knowledge of the explicit form of the unitary transformation which generates the target state. As a result, that approach does not offer any advantages for real-world applications. In recent experimental works, VQAs have shown potential for preparing some specific mixed quantum states, such as thermal Gibbs states at various temperatures in a system of two transmon qubits \cite{Sagastizabal_2021}.

In this work, we focus on the experimental variational preparation of two-qubit Bell states and the three-qubit Greenberger–Horne–Zeilinger (GHZ) state \cite{Nielsen_Chuang_2010} using the superconducting quantum computing platform. Typically, Bell states are generated using high-precision two-qubit entangling gates, such as $\sqrt{\text{iSwap}}$ \cite{Steffen_2006, Ansmann_2009} or cPhase \cite{DiCarlo_2009}. Obviously, the fidelity of the resulting states is highly sensitive to the accuracy of the gates, which in turn demands intricate calibration procedures \cite{Arute_2020}. To show that this challenge may be addressed, in this work we develop a fully automatic calibration routine based on a gradient-descent VQA that incorporates a quantum circuit comprising minimally calibrated fixed iSwap-like gates \cite{Foxen_2020} and single-qubit $X$ and $Y$ rotations with tunable angles. Our approach introduces an alternative methodology for learning accurate quantum evolution using non-ideal multi-qubit gates, and, simultaneously, exhibits a clear example of a practical application for variational quantum algorithms in realistic experimental settings. 

The minimally calibrated iSwap-like gate that we use emerges physically from the multi-level structure of transmons leading to an evolution combining the $|01\rangle \leftrightarrow |10\rangle$ iSwap interaction with residual phase shifts from unintended $|11\rangle \leftrightarrow |02\rangle$ cPhase coupling; additionally, single-qubit phases are accumulated. While in other applications such as digital quantum algorithms this kind of behavior may be regarded as disadvantageous, our approach instead leverages this native operation by compensating for its imperfections through variational optimization of surrounding single-qubit gates, reproducing high-fidelity state preparation. 

We confirm the fidelity of the generated entangled states, by performing the quantum state tomography (QST) \cite{Liu_2005}. However, full QST is not necessary for entanglement verification, since Bell states exhibit strong correlations in measurements which may be revealed in experiments requiring only single-qubit rotations and simultaneous measurement of qubit states. In 1964, Bell formulated an inequality for two entangled particles \cite{Bell_1964}, providing a framework to experimentally test whether quantum mechanics is the best possible theory or whether there exist some underlying hidden local variables knowing which a better theory without uncertainty could be developed \cite{Einstein_1935}. The Clauser–Horne–Shimony–Holt (CHSH) inequality, derived in 1969 \cite{Clauser_1969}, simplified the experimental verification of Bell's theorem. We experimentally measured the violation of the CHSH inequality for the prepared Bell states, demonstrating their entanglement and nonclassical behavior.

\section{\label{sec:level1}Experimental design}
\label{section: experimental design}

\begin{figure*}[t]
\includegraphics[width=1\linewidth]{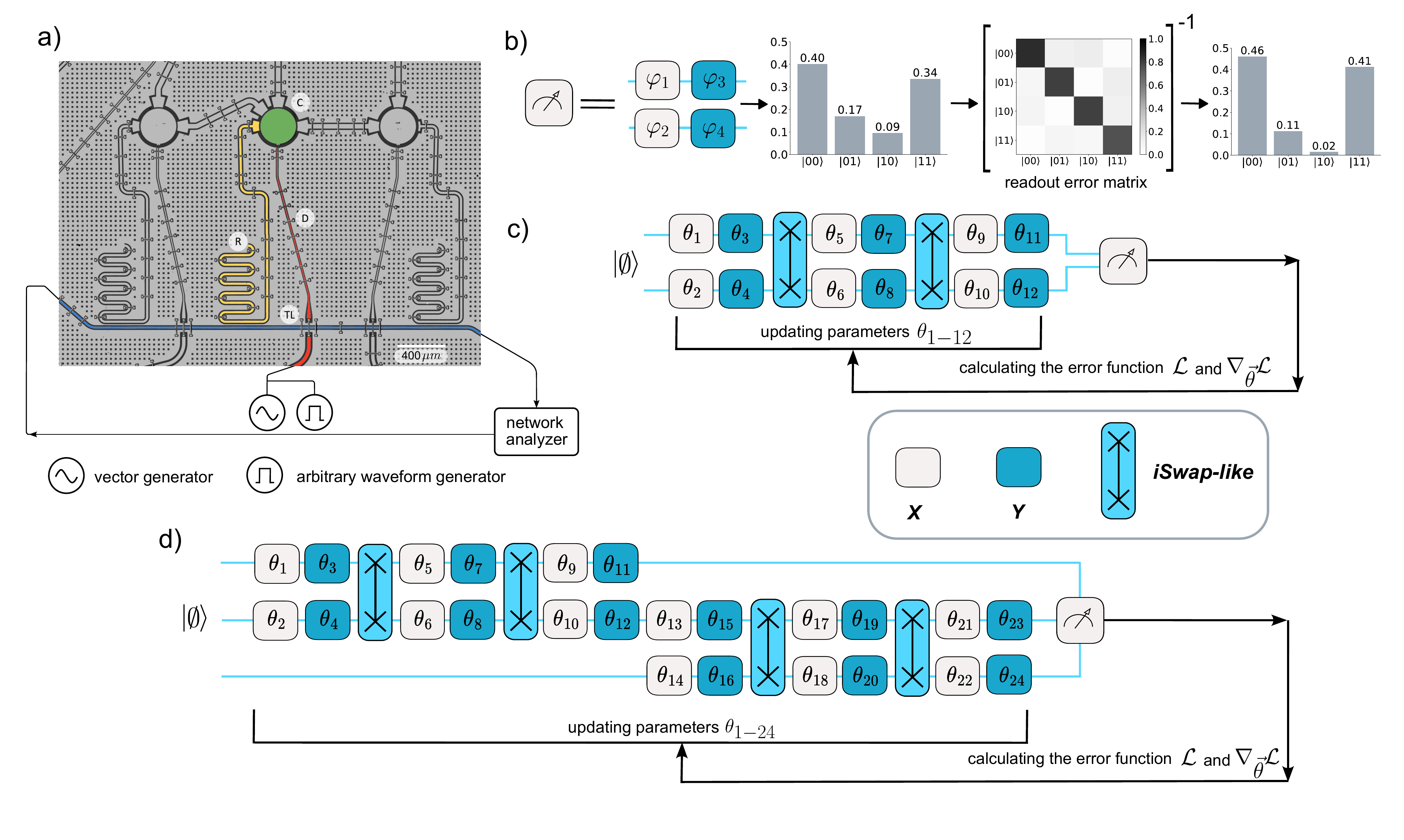}% Here is how to import EPS art
\caption{
a) Micrograph of the three transmons used in this experiment (false-colored). Green (C) represents the transmon shunt capacitance, yellow (R) denotes the readout resonator, red (D) indicates the drive-bias control line, and blue (TL) corresponds to the transmission readout line. b) The state readout protocol involves three sequential steps: (1) tomography rotations to prepare various measurement bases, (2) statistical averaging through repeated measurements (about 2000 shots) to determine basis state populations, and (3) application of readout error correction using the inverse error matrix method. c) Quantum circuit for preparing Bell states with two qubits. d) Quantum circuit for preparing GHZ states with three qubits.
}
\label{fig:chip_quantum_chain}
\end{figure*}

The experiment was conducted on three transmon qubits \cite{Koch_2007} (I, II, and III, left to right) as shown in Figure~\ref{fig:chip_quantum_chain}a. The entire experimental sample incorporates 16 superconducting transmon artificial atoms (see Supplementary for full device characterization). The key properties of the studied transmons, including transition frequencies ($\nu_{01}$), relaxation times ($T_1$), dephasing times ($T_2^{*}$), and their readout resonator frequencies ($\nu_r$), are detailed in Table~\ref{tab:qubits}. Experimental equipment, the measurement scheme is almost the same as described in the article \cite{Tolstobrov_2024_RF}.

As said above, we leverage the native iSwap-like gates realizing a population exchange between the $|01\rangle$ and $|10\rangle$ states in a pair of qubits. These gates are roughly calibrated by varying the amplitude and duration (with 1 ns resolution) of the flux pulse applied to one of the transmons and finding an optimum on a two-dimensional map of population transfer (data in Supplementary). To perform high-quality single-qubit rotations, we implemented the DRAG (Derivative Removal by Adiabatic Gate) calibration scheme \cite{Motzoi_2009, Gambetta_2011}, which significantly mitigates leakage to higher transmon energy levels. Using single-qubit randomized benchmarking \cite{Knill_2008}, we measured an average gate fidelity of $99.74\%$ for single-qubit rotations of angles $\pi/2$ and $\pi$, confirming high-precision control of individual qubits in our system.

The transmons are read out using a frequency-multiplexed scheme \cite{Chen_2012} in single-shot mode \cite{Walter2017}, enabling the measurement of multi-qubit correlations. We use a Josephson parametric amplifier (JPA) \cite{Dorogov2022} to enhance the signal-to-noise ratio for the single-shot readout. Given the resonator frequency separation of approximately 300–400 MHz between resonators neighboring qubits (see Table \ref{tab:qubits}), the readout accuracy achieved with a narrowband JPA was approximately 80–85\%.  To mitigate readout errors, we applied the inverse error matrix method \cite{Maciejewski_2020}, where the error matrix was measured directly after preparing the qubits in their basis states.

\begin{table}[h]
\caption{\label{tab:qubits}%
The measured parameters for the superconducting artificial atoms: $\nu_{01}$ - the frequency of the transition from the ground state \(|0\rangle\) to the excited state \(|1\rangle\), $T_1$ - the qubit relaxation time, characterizing energy decay, $T_2^{*}$ - the qubit dephasing time, representing coherence loss, $\nu_r$ - the frequency of the resonator coupled to the transmon, and durations of two-qubit operation (iSwap-like gates) and single-qubit rotations.
}
\begin{ruledtabular}
\begin{tabular}{l c c c }
\textrm{Qubit} & \textrm{I} & \textrm{II} & \textrm{III} \\
\colrule
$sweet\ spot$        & $bottom$     & $top$       & $bottom$     \\
$\nu_{01}, \text{GHz}$      & 4.228        & 4.747       & 4.497        \\
$T_1, \mu \text{s}$         & 22           & 16          & 23           \\
$T_2^{*}, \mu \text{s}$         & 3.5          & 3.2         & 4.8          \\
$\nu_{r}, \text{GHz}$       & 6.717        & 6.436       & 6.827        \\
$iSwap,\text{ns}$          & \multicolumn{3}{c}{37\ \ \ \ \ \ \ \ \ \ \ \ \ \ \ \ \ 26}\\
$X, Y, \text{ns}$           & 40           & 40          & 40           \\
\end{tabular}
\end{ruledtabular}
\end{table}

\section{\label{sec:level1}Preparation of Bell's states and GHZ state}

Figure \ref{fig:chip_quantum_chain}b,c illustrates the quantum circuit designed for preparing the two-qubit Bell states. The circuit comprises 12 parameterized single-qubit rotations, 6 around the $X$-axis and 6 around the $Y$-axis of the Bloch sphere, along with two non-ideal iSwap-like entangling operations. Following the state preparation block, the circuit includes a module for quantum state tomography and qubit measurement.

We begin the optimization procedure for generating the Bell states with preparation of the multi-qubit ground state $\left|{\stackinset{c}{}{c}{}{/}{0}}\right\rangle$. It is achieved through qubit relaxation, with a waiting time of approximately $5\,T_1$ to ensure high-fidelity initialization; however, this does not reduce the presence of thermal population of the qubit \cite{lvov2025thermometrybasedsuperconductingqubit}. Alternatively, active reset algorithms \cite{Geerlings_2013, Magnard_2018} can be employed to significantly reduce the initialization time and accelerate the optimization process. 

After the variational ansatz is applied, the loss function needs to be calculated. For the parameters vector $\bm{\theta} = \theta_{1-12}$ (at the first step of the algorithm it is determined by random values) of the single-qubit $X,\ Y$ rotations, the probabilities of measuring the qubits in each of the four basis states $|00\rangle, \ |01\rangle,\ |10\rangle, \ |11\rangle$ are determined, see Figure \ref{fig:chip_quantum_chain}b. For this purpose, we perform 2000 runs of the quantum circuit execution and average the results. 

The loss function is then calculated as:
\begin{equation}
\mathcal{L}\ =\frac{1}{4N} \sum_{i=1}^{N}\sum_{j=1}^{4} (p_\text{targ}^{(i,j)}-p_\text{exp}^{(i,j)})^2,
\label{eq:loss}
\end{equation}
where $j$ denotes the indices of the measured states, $i$ represents the indices of the pre-measurement rotations: $X_{\varphi_{1}},X_{\varphi_{2}},Y_{\varphi_{3}},Y_{\varphi_{4}} = \{[\mathbb{1},\mathbb{1}],[\mathbb{1},X_{\frac{\pi}{2}}],[X_{\frac{\pi}{2}},\mathbb{1}]\}\otimes\{[\mathbb{1},\mathbb{1}],[\mathbb{1},Y_{\frac{\pi}{2}}],[Y_{\frac{\pi}{2}},\mathbb{1}]\}$, $N$ is number of all combinations of tomography rotations; finally, $p_\text{targ}^{(i,j)}$ are the theoretical probabilities for the target state and $p_\text{exp}^{(i,j)}$ are the experimentally measured probabilities for the current set of parameters $\bm \theta$, including readout error correction. For example, for the Bell state $(|00\rangle + |11\rangle)/\sqrt{2}$ and rotations $\varphi_{1-4}^{(i=1)} =0$ one has $p_\text{targ}^{(i=1, j=1:4)}= [0.5,\ 0,\ 0,\ 0.5] $. Usually, for a two-qubit system, quantum state tomography requires 15 distinct measurement bases with measurement of the correlator of qubit states to reconstruct the density matrix $\rho$. However, our protocol requires just $N = 9$ distinct measurement bases because we read out the populations of all basis states, which carries more information. Due to the normalization constraint, each basis yields 3 independent probability measurements, resulting in 27 total parameters for the loss function $\mathcal{L}$. This exceeds the 15 independent parameters needed to reconstruct the density matrix $\rho$, ensuring the generated state can be completely determined. 

Finally, the protocol completes with updating the ansatz parameters $\bm\theta$. The parameters of the quantum circuit are optimized using Nesterov's accelerated gradient descent algorithm \cite{Nesterov1983AMF}. The gradients of the loss function with respect to the parameters are computed using the parameter-shift rule \cite{Mitarai2018, Schuld2018EvaluatingAG}, which enables efficient gradient estimation for variational quantum algorithms. 

After the convergence is reached, we perform the usual QST for the Bell states based on the measurement results for all rotation angles combinations $\varphi^{(i)}_{1-4}$ via maximum likelihood estimation. The density matrix \(\rho\) is parameterized using the Choletsky decomposition. We calculated the measurement probabilities for current parametrization of density matrix using the cross-platform Python library PennyLane \cite{PennyLane_2022} for all tomography angles, and then the mean square of deviation from measured probabilities was calculated and minimized to reconstruct the density matrix.

\begin{figure}[t]
\includegraphics[width=1\linewidth]{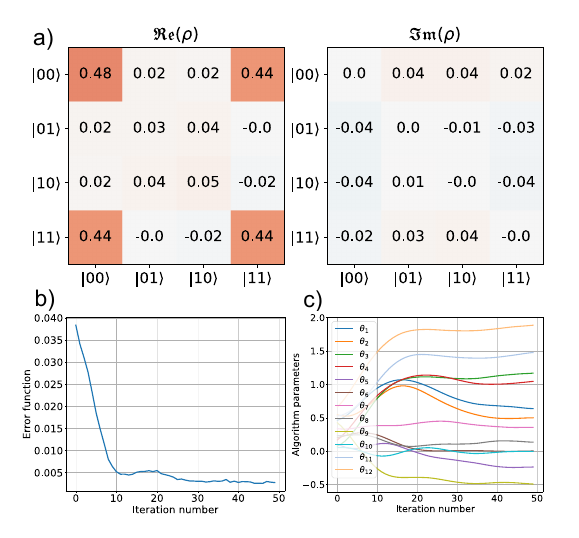}
\hfill
\caption{a) For the Bell state $\frac{1}{\sqrt{2}}(|00\rangle + |11\rangle)$, the figure presents the experimentally measured density matrix at the last iteration of the algorithm, showing both real and imaginary components. b) The evolution of the loss function $\mathcal{L}$ as a function of algorithm iteration number, demonstrating the convergence behavior. c) The parameters optimization trajectory, showing the dependence of quantum circuit parameters on the algorithm iteration number.}
\label{ris:bell satates}
\end{figure}

Figure \ref{ris:bell satates} presents the optimization data for one of the Bell states $|\beta_{00}\rangle = (|00\rangle + |11\rangle)/\sqrt{2}$: density matrices of the two-qubit states obtained from quantum state tomography (showing both real and imaginary parts), the dependence of the loss function \(\mathcal{L}\) on the algorithm iteration, the evolution of the parameters \(\theta_{1-12}\) as a function of the algorithm iteration number.

The Supplementary information presents complete experimental results for all four Bell states. The fidelities $\mathcal{F}=\sqrt{\text{Tr}(\rho_{exp}\cdot\rho_{targ})}$ ($\rho_{exp}$ is the reconstructed density matrix of the prepared state, $\rho_{targ}$ is the density matrix of the target state) of the density matrices, along with the standard deviations computed as the average over the last five steps of the algorithm, are summarized in Table \ref{tab:fidelities}.

\begin{table}[h]
\renewcommand{\arraystretch}{2.0}
\begin{ruledtabular}
\begin{tabular}{c c c}
\textrm{Bell State} & {\textrm{Fidelity $\mathcal{F} \pm \sigma$}} & $max(|S_{1,2}|)\pm \sigma$\\
\colrule
$|\beta_{00}\rangle = \dfrac{1}{\sqrt{2}}(|00\rangle + |11\rangle)$ & 0.949 $\pm$ 0.005 &  2.47 $\pm$ 0.08 \\ 
$|\beta_{01}\rangle = \dfrac{1}{\sqrt{2}}(|01\rangle + |10\rangle)$ & 0.987 $\pm$ 0.009 &
2.77 $\pm$ 0.10\\
$|\beta_{10}\rangle = \dfrac{1}{\sqrt{2}}(|00\rangle - |11\rangle)$ & 0.930 $\pm$ 0.006 &
2.30 $\pm$ 0.11\\
$|\beta_{11}\rangle = \dfrac{1}{\sqrt{2}}(|01\rangle - |10\rangle)$ & 0.956 $\pm$ 0.009 &
2.52 $\pm$ 0.11\\
\end{tabular}
\end{ruledtabular}
\caption{\label{tab:fidelities} Fidelities of generated Bell states with standard deviations and maximum values of CHSH inequality violation with standard deviations.}
\end{table}

The next application of the variational quantum algorithm (VQA) is the preparation of the Greenberger–Horne–Zeilinger (GHZ) state: $(|000\rangle + |111\rangle)/{\sqrt{2}}$.
This three-qubit entangled state represents a generalization of the Bell state to three qubits. The quantum circuit used for its preparation, shown in Figure \ref{fig:chip_quantum_chain}d, extends the Bell state circuit incorporating 12 single-qubit rotations around the $X$-axis of the Bloch sphere, 12 single-qubit rotations around the $Y$-axis, and four native iSwap-like entangling operations. 

\begin{figure}[t]
\includegraphics[width=0.8\linewidth]{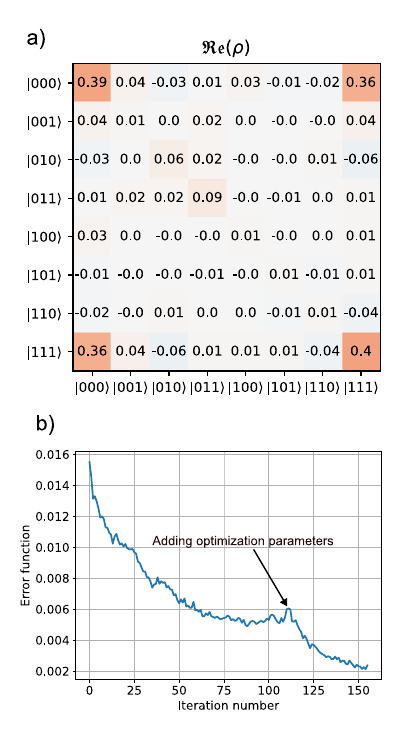}% Here is how to import EPS art
\caption{\label{fig:000+111} a)The real component of the reconstructed density matrix of the three-qubit GHZ state $\frac{1}{\sqrt{2}}(|000\rangle + |111\rangle)$ obtained after variational circuit optimization. 
b)The values of the error function throughout the algorithm iterations are shown. The training procedure was initialized with the state \(\frac{1}{\sqrt{2}}(|00\rangle + |11\rangle)\), first optimizing parameters \(\theta_{13-24}\). The arrow indicates the transition point where optimization of all 24 parameters (\(\theta_{1-24}\)) commenced.
}
\end{figure}

The optimization process and density matrix tomography follow the same approach as described for the two-qubit case, generalized to three qubits. Since for a 3-qubit state the calculation of the loss function $\mathcal{L}$ takes more time, the optimization was carried out as follows: the Bell state $\frac{1}{\sqrt{2}}(|00\rangle + |11\rangle)$, already obtained through optimization, was used as input, and parameters $\theta_{13-24}$ were optimized. After reaching convergence, we included all 24 parameters $\theta_{1-24}$ in the optimization. Figure~\ref{fig:000+111} displays the real part of the reconstructed density matrix for the GHZ state obtained through variational circuit optimization  and the values of the loss function throughout the algorithm iterations. The fidelity of the prepared state, calculated from the density matrix, is $0.869 \pm 0.003$. Further details about the GHZ state preparation are provided in the Supplementary Materials.

\section{\label{sec:level1}Demonstration of CHSH inequality violations}

We employ two combinations of correlators to measure the CHSH inequality:

\begin{equation}
S_1=E(a,b)+E\left(a',b\right)+E\left(a,b'\right)-E\left(a',b'\right), 
\label{eq:S_1}
\end{equation}
and
\begin{equation}
S_2=-E(a,b)-E\left(a',b\right)+E\left(a,b'\right)-E\left(a',b'\right), 
\label{eq:S_2}
\end{equation}
where $E(a,b) = P_{a,b}(|00\rangle) - P_{a,b}(|10\rangle) - P_{a,b}(|01\rangle) + P_{a,b}(|11\rangle)$ is the correlator of measured states for two qubits at rotation angles $a$ and $b$, and $P$ denotes the probability of measuring one of the basis states for the two-qubit system.

The combination $S_1$ is used for the Bell states $|\beta _{00}\rangle =(|00\rangle +|11\rangle )/\sqrt{2}$ and $|\beta _{01}\rangle = (|01\rangle +|10\rangle )/\sqrt{2}$, while $S_2$ is applied to the states $|\beta _{10}\rangle =(|00\rangle -|11\rangle )/\sqrt{2}$ and $|\beta _{11}\rangle = (|01\rangle -|10\rangle )/ \sqrt{2}$. For entangled states, the violation of the CHSH inequality requires $|max(|S_1|)|>2$ and $|max(|S_2|)|>2$, with the maximum theoretical value bounded by $2\sqrt{2}$ (Tsirelson's bound) \cite{Cirelson_1980}.

\begin{figure}[t]
\includegraphics[width=0.8\linewidth]{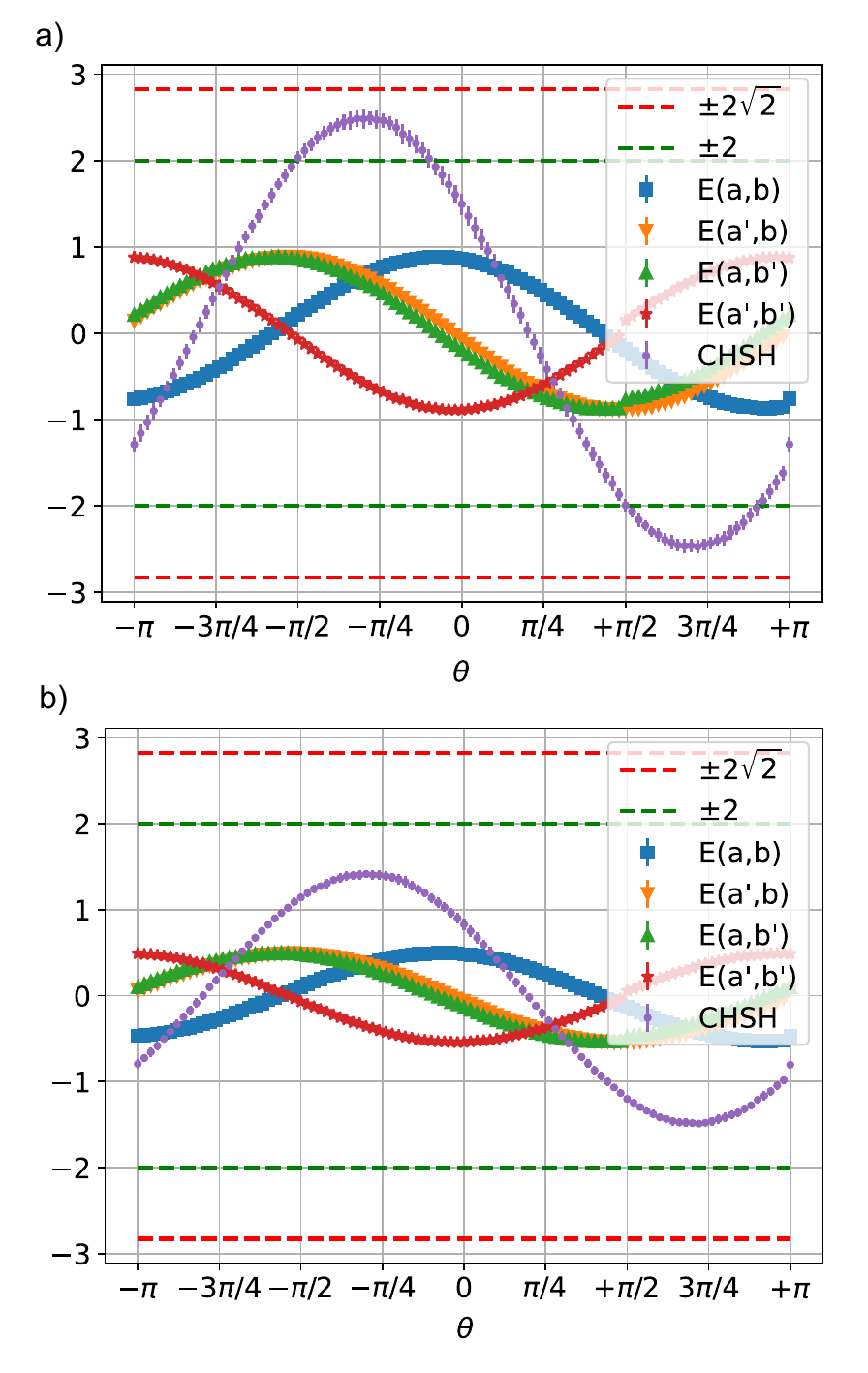}% Here is how to import EPS art
\caption{\label{fig:correction}Correlator and CHSH values for the Bell state $\frac{1}{\sqrt{2}}(|00\rangle + |11\rangle)$ with readout error correction a) and without correction b).}
\end{figure}

Figure \ref{fig:correction}a presents the experimental measurements of the CHSH inequality violation. The rotation angles of the qubits around the $X$-axis of the Bloch sphere are defined as:
$$
a' = a + \pi/2, \quad b' = b + \pi/2, \quad \theta = a - b.
$$
The graph for one of the Bell states $|\beta_{00}\rangle = (|00\rangle + |11\rangle)/ \sqrt{2}$ depicts the dependence of the correlators $E(a,b),\ E(a',b),\ E(a,b')$, and $E(a',b')$, as well as the expressions $S_{1}$, on the angle $\theta$. Each point is the result of averaging over 50 experiments; sticks show standard deviations. The Supplementary information presents CHSH inequality violation measurements for all four Bell states. The results demonstrate that the maximum values of \(|S_{1,2}|\) exceed the classical limit of 2 for all Bell states: \(|\beta_{00}\rangle\), \(|\beta_{01}\rangle\), \(|\beta_{10}\rangle\), and \(|\beta_{11}\rangle\). Specifically,the measured maxima are summarized in Table \ref{tab:fidelities}.

At angles \(\theta = \pi/2\) and \(\pi\), notches in the correlators are observed. These arise due to the transition between \(R_x(-\pi)\) and \(R_x(\pi)\) rotations, which is sensitive to non-ideal calibration of the rotation angles, despite the high precision of the estimation according to single-qubit randomized benchmarking (see Section \ref{section: experimental design}).

% \begin{figure}[h]
% {\includegraphics[width=1\linewidth]{CHSH.pdf}}
% % \begin{minipage}[h]{0.49\linewidth}
% % \center{\includegraphics[width=0.78\linewidth]{CHSH.png} \\ a)}
% % \end{minipage}
% % \hfill
% % \begin{minipage}[h]{0.49\linewidth}
% % \center{\includegraphics[width=0.45\linewidth]{correction.png} \\ b)}
% % \end{minipage}
% \caption{Violation of the CHSH inequality for 4 Bell states, the value of the inequality is above the classical limit of 2.}
% \label{ris:CHSH}
% \end{figure}

Figure \ref{fig:correction}b shows the measurements of the correlators without readout error correction. More details on the error correction procedure are provided in Section \ref{section: experimental design}. The results clearly demonstrate that, in the absence of readour correction, the CHSH inequality is not violated due to the significant impact of readout errors.

\section{\label{sec: Conclusions} Conclusions}

This work demonstrates the application of variational quantum algorithm (VQA) to prepare entangled Bell states and Greenberger–Horne–Zeilinger (GHZ) states. By utilizing a simple variational circuit incorporating non-ideal iSwap-like gates and single-qubit $X$ and $Y$ rotations, we achieved high-fidelity preparation of these states. The average fidelity of the Bell's prepared states is $0.956 \pm 0.007$. For the GHZ state, the fidelity is $0.869 \pm 0.003$. The key advantages of our approach lie in the simplicity of the circuit design with the minimal calibration required for the iSwap-like gates and ideal quantum evolution can be precisely reproduced using a variational circuit containing imperfect two-qubit gates.

Additionally, we experimentally measured the violation of the Clauser–Horne–Shimony–Holt (CHSH) inequality for the prepared Bell states, confirming their quantum entanglement. The observed violations exceeded the classical limit of 2, with average value $2.52 \pm 0.10$.

However, scaling this direct approach to larger qubit systems becomes challenging due to the exponentially increasing tomography time for multi-qubit states. To address this, a hybrid strategy can be employed: a standard circuit with Hadamard and CNOT gates can be used for state preparation, while the approximation of two-qubit gates can be achieved using variational circuits similar to those developed in this work.

\section*{\label{sec: Conclusions} Acknowledgments}
The experimental sample was fabricated at the MIPT Shared Facility Center. All authors declare no competing interests.

% The \nocite command causes all entries in a bibliography to be printed out
% whether or not they are actually referenced in the text. This is appropriate
% for the sample file to show the different styles of references, but authors
% most likely will not want to use it.
\nocite{*}

\bibliography{bibliography}% Produces the bibliography via BibTeX.

\end{document}

	% \preprint{AIP/123-QED}
	
	\title[\mytitile]{\mytitile\\~}
	
\maketitle

\section{Experimental sample}

This section provides an overview of the experimental sample used in the study. The illustrations include:
\begin{itemize}
    \item A 16-transmon quantum processor, with a focus on the three qubits involved in the experiment (highlighted in Figure 1a).
    \item A micrograph of the chip bonded into a holder (Figure 1b).
\end{itemize}

\begin{figure}[h]
\includegraphics[width=1.0\linewidth]{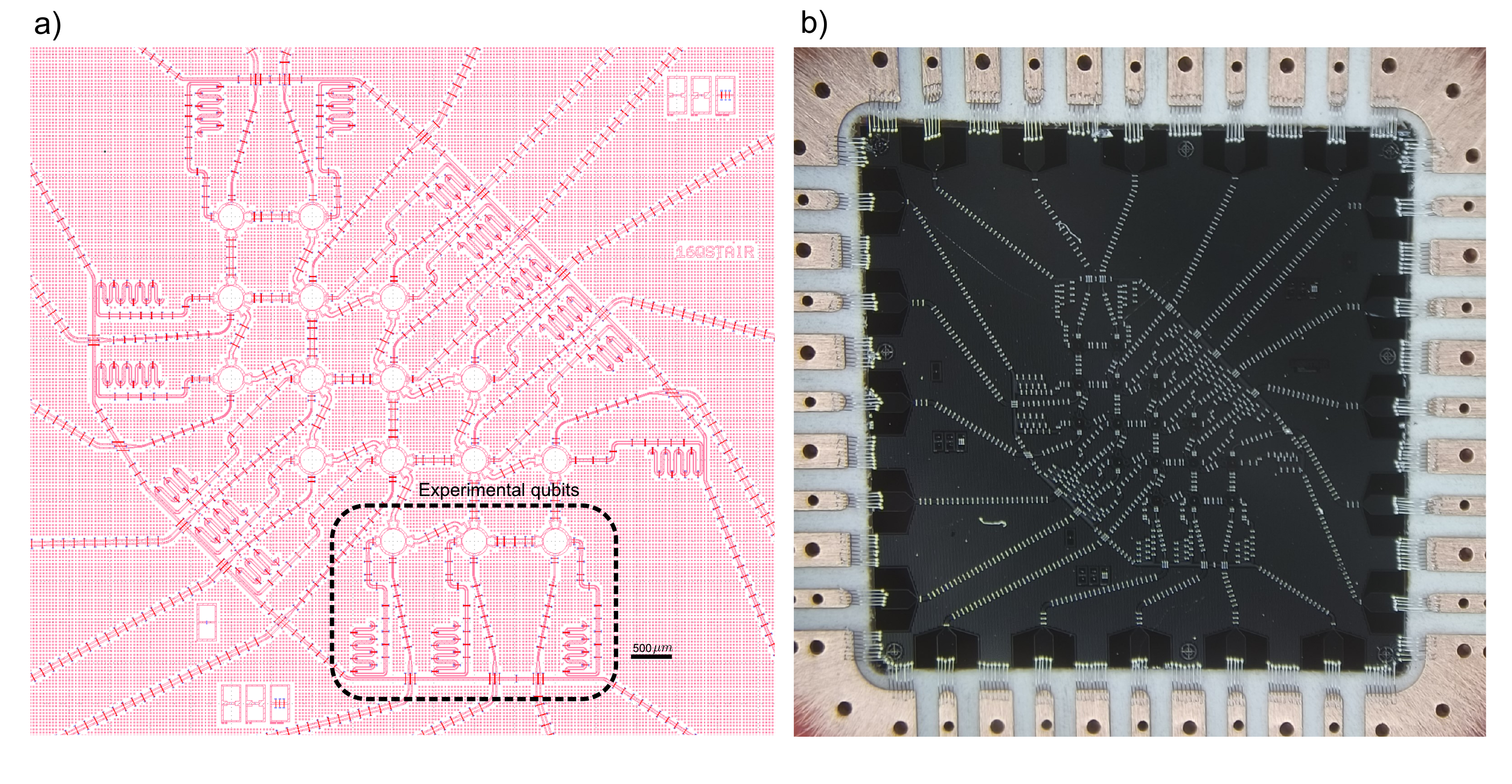}% Here is how to import EPS art
\caption{\label{fig:general}a) Schematic of the 16-transmon quantum processor. The three qubits used in the experiment are highlighted within the dotted-line boundary. b) Micrograph of the chip bonded into a holder.}
\end{figure}

\section{Calibration of iSwap-like two-qubit gate} 
\label{sec: calibration of iSwap-like two-qubit gate}

\begin{figure}[h]
\includegraphics[width=0.5\linewidth]{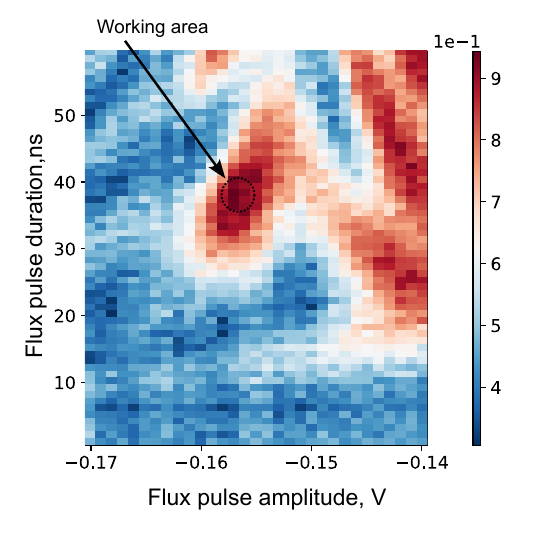}% Here is how to import EPS art
\caption{\label{fig:iSwap} iSwap-like gate calibration. The arrow shows the area of maximum transfer of populations between the $|01\rangle,\ |10\rangle$ transmon levels.}
\end{figure}

The calibration of the two-qubit iSwap-like operation was performed as follows. We excite one of the qubits from level $|0\rangle$ to $|1\rangle$ and drive two qubits into resonance to observe oscillations between $|01\rangle$ and $|10\rangle$ qubit states, which can be seen in Figure \ref{fig:iSwap}. Thus using two parameters, the operation time and the amplitude of the flux pulse, we find the point of maximum population transfer between states $|01\rangle$, $|10\rangle$, as depicted in Figure 2.

\section{Tomography of iSwap-like two-qubit gate}
\label{sec: tomography of two-qubit gates}

Quantum process tomography \cite{Liu_2005} was used to understand the appearance of simple calibrated two-qubit iSwap-like gates. Then, a two-qubit gait approximation \cite{Foxen_2020} was performed to best fit the resulting process matrices. 

\begin{figure}[h]
\includegraphics[width=1.0\linewidth]{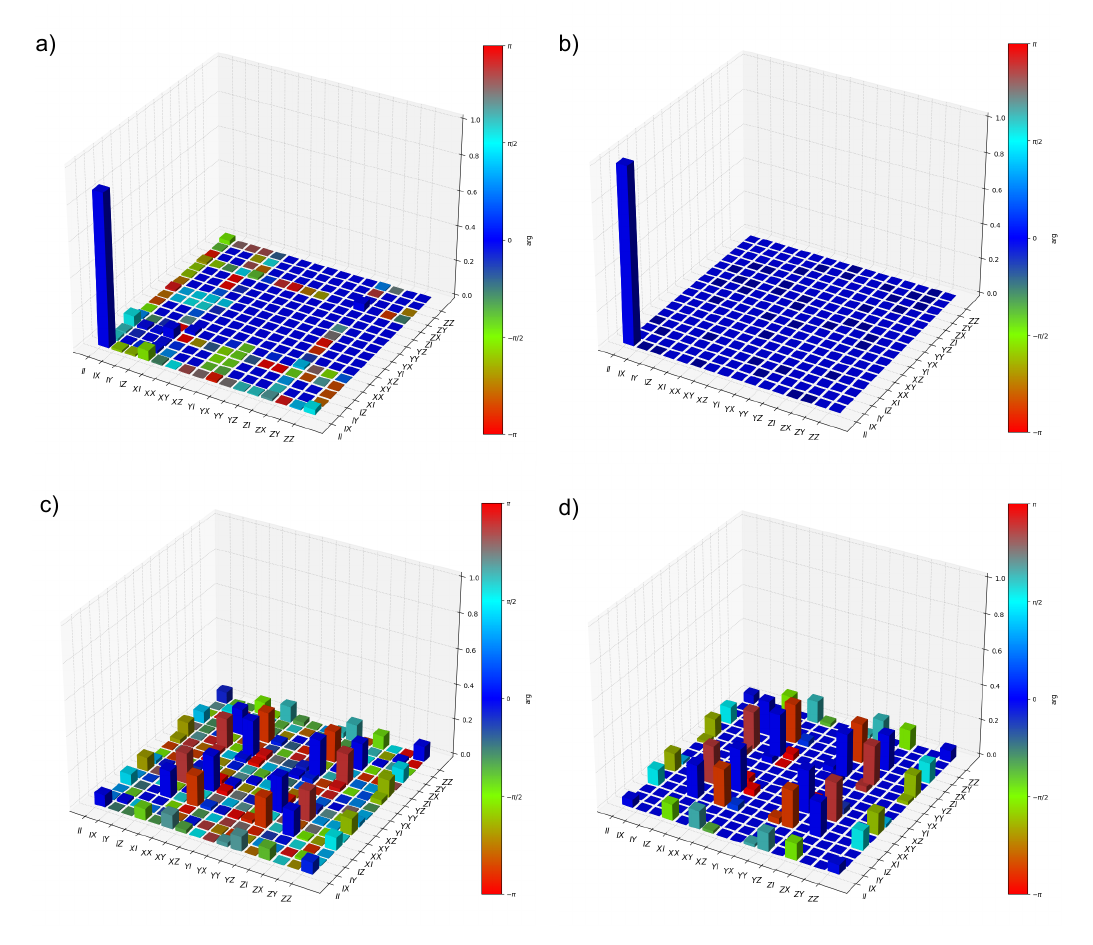}% Here is how to import EPS art
\caption{\label{fig:tomography}a)-b) Tomography of the identity operation. Fidelity of tomography: 0.93. c)-d) Tomography of the two-qubit iSwap-like operation and approximation using \ref{eq:fsim} matrix. Obtained parameter values: $\theta = 1.52 $ , $\phi = 1.21 $, $\Delta_{+} = -1.69 $, $\Delta_{-} = 0.41 $, $\Delta_{-, o f f} = 0.15 $. Fidelity of the approximation of the process matrix: 0.89.}
% c)-d) Tomography of the two-qubit iSwap-like operation between XV-XVI qubits and approximation. Obtained parameter values: $\theta = 1.51 $ , $\phi = 1.02 $, $\Delta_{+} = -0.55 $, $\Delta_{-} = 0.85 $, $\Delta_{-, o f f} = 1.69 $. Fidelity of the approximation of the process matrix: 0.90.
\end{figure}

We approximate the matrix of a two-qubit quantum operation by the \ref{eq:fsim} model. Obtained parameter values for iSwap-like operation: $\theta = 1.52 $ , $\phi = 1.21 $, $\Delta_{+} = -1.69 $, $\Delta_{-} = 0.41 $, $\Delta_{-, o f f} = 0.15 $. The fidelity we define as $\mathcal{F}=\sqrt{\text{Tr}(M_{exp}\cdot M_{targ})}$ ($M_{exp}$ is the measured matrix of the quantum process, $M_{targ}$ is the target matrix of the quantum process) of the two-qubit gate (see Figure \ref{fig:tomography}). Fidelity of the approximation of the process matrix: $\mathcal{F}_{final}=0.89$. Fidelity of the identity operation: $\mathcal{F}_{id}=0.93$. To estimate the effect of reading errors on tomography, it will be assumed that the resulting fidelity is equal to $\mathcal{F}_{final}=\mathcal{F}_{id}\cdot\mathcal{F}_{iSwap-like}$. Thus the approximation fidelity of the iSwap-like operation is equal to $\mathcal{F}_{iSwap-like} = 0.96$.

% Obtained parameter values for iSwap-like operation between XV-XVI qibits: $\theta = 1.51 $ , $\phi = 1.02 $, $\Delta_{+} = -0.55 $, $\Delta_{-} = 0.85 $, $\Delta_{-, o f f} = 1.69 $. Fidelity of the approximation of the process matrix: 0.90.

It can be seen that iSwap parameter $\theta$ is close to $\pi/2$, which is in agreement with our simple calibration. But the iSwap-like operation also has a phase from the cPhase operation and single-qubit phases.

\begin{equation}
\begin{aligned}
& \operatorname{iSwap-like}\left(\theta, \phi, \Delta_{+}, \Delta_{-}, \Delta_{-, o f f}\right)= \\
& \qquad\left(\begin{array}{cccc}
1 & 0 & 0 & 0 \\
0 & e^{i\left(\Delta_{+}+\Delta_{-}\right)} \cos \theta & -i e^{i\left(\Delta_{+}-\Delta_{-, o f f}\right)} \sin \theta & 0 \\
0 & -i e^{i\left(\Delta_{+}+\Delta_{-, o f f}\right)} \sin \theta & e^{i\left(\Delta_{+}-\Delta_{-}\right)} \cos \theta & 0 \\
0 & 0 & 0 & e^{i\left(2 \Delta_{+}+\phi\right)}
\end{array}\right)
\end{aligned}
\label{eq:fsim}
\end{equation}

\section{Results of preparation of Bell states and GHZ state}

\begin{figure}[h]
\includegraphics[width=0.7\linewidth]{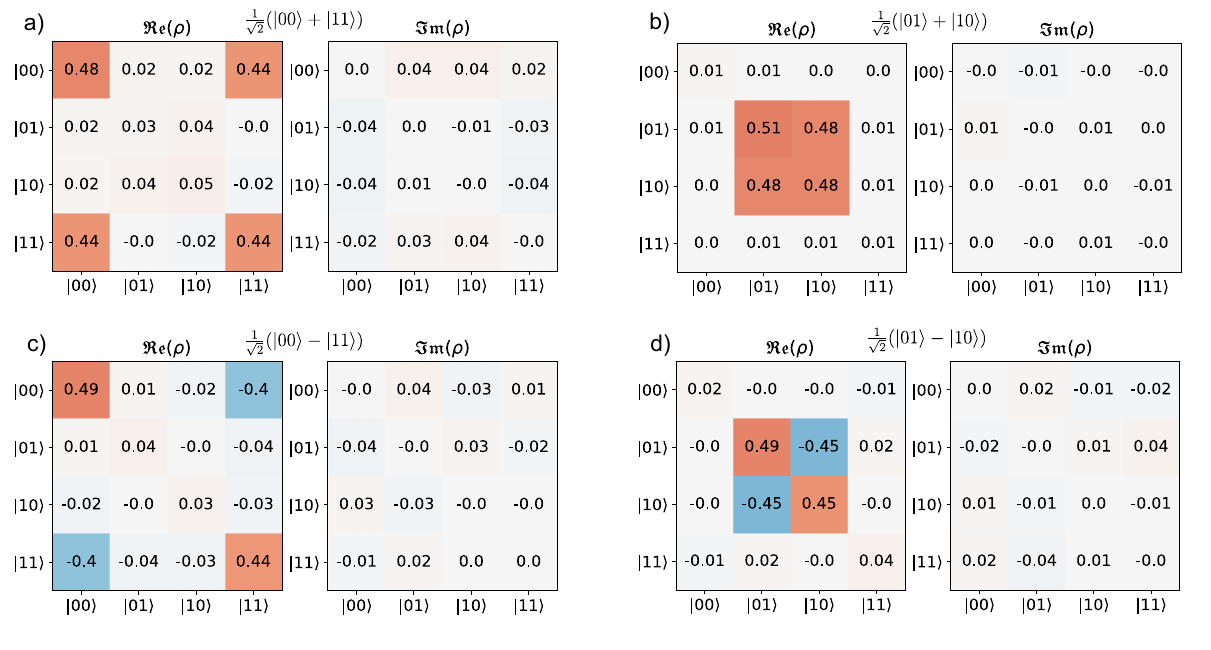}% Here is how to import EPS art
\caption{\label{fig:bell_states} Measured density matrices, their real and imaginary parts, for 4 Bell states.}
\end{figure}

\begin{figure}[h]
\includegraphics[width=0.65\linewidth]{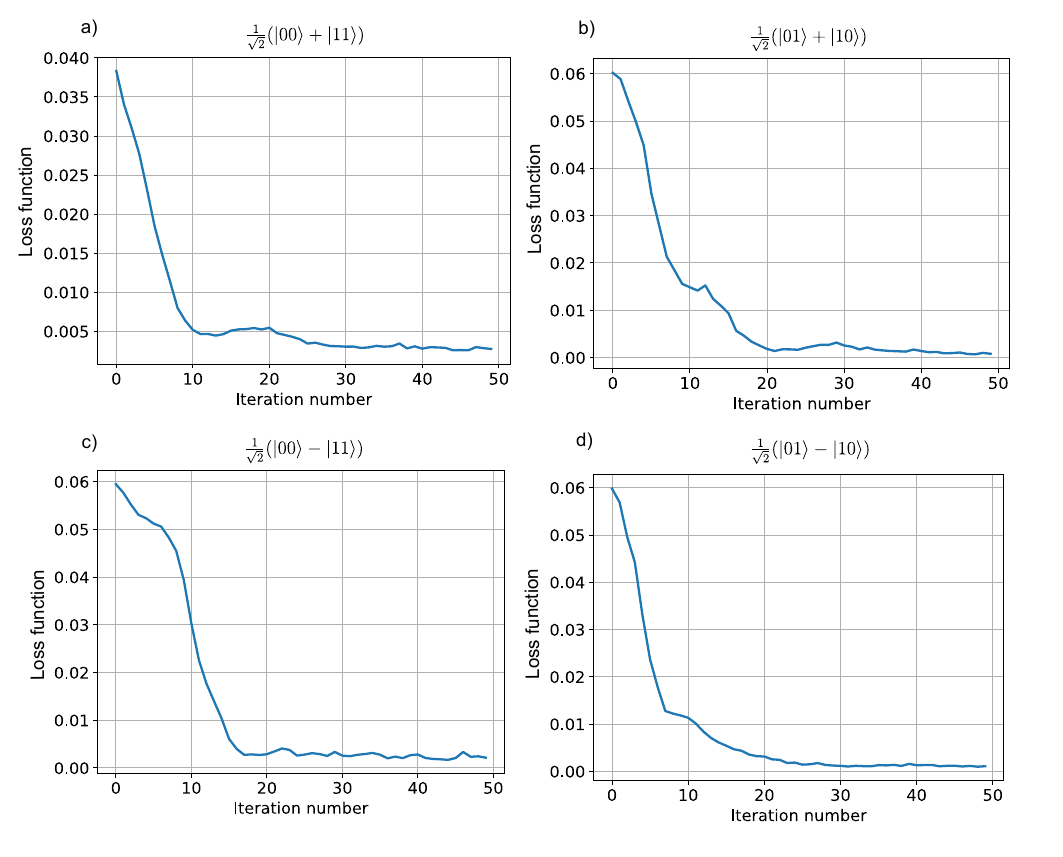}% Here is how to import EPS art
\caption{\label{fig:costs} Dependencies of the loss function on the number of iterations of the optimization algorithm for 4 Bell states.}
\end{figure}

\begin{figure}[h]
\includegraphics[width=0.65\linewidth]{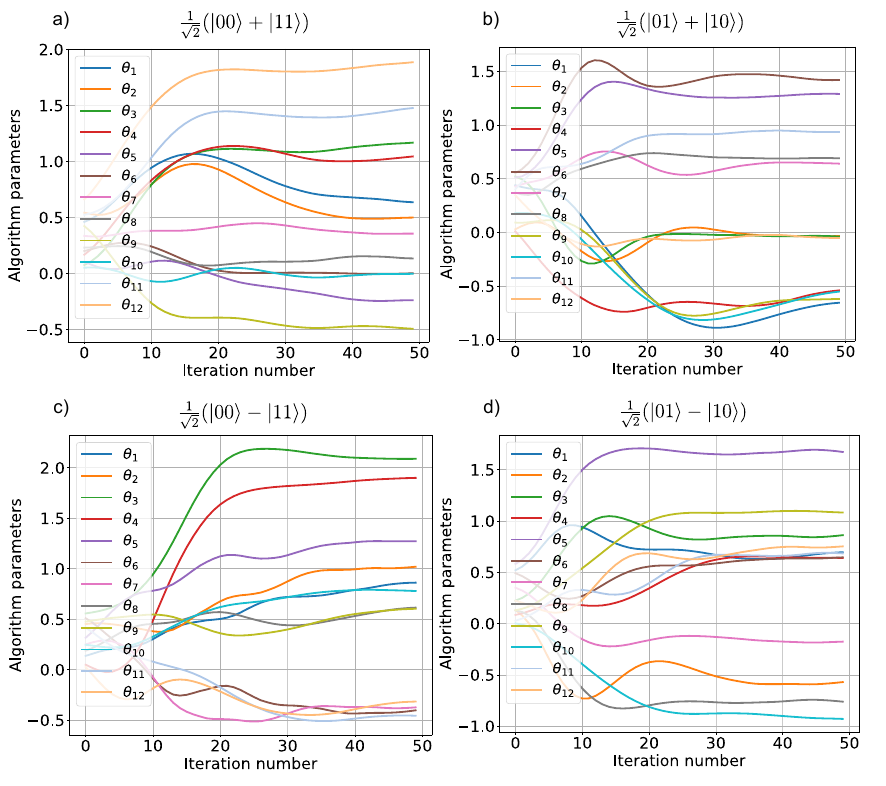}% Here is how to import EPS art
\caption{\label{fig:params} Evolution of the optimized parameters of the algorithm from the iteration number for 4 Bell states.}
\end{figure}

\begin{figure}[h]
\includegraphics[width=0.6\linewidth]{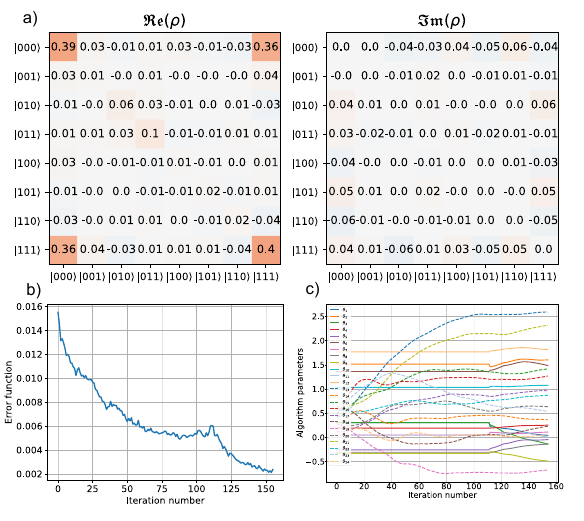}% Here is how to import EPS art
\caption{\label{fig:000+111} a) Measured density matrix for three-qubit GHZ state. b) Dependencies of the loss function on the number of iterations of the optimization algorithm for GHZ state. c) Evolution of the optimized parameters of the algorithm from the iteration number for GHZ state.}
\end{figure}

\clearpage
\section{Results of measuring of CHSH inequality violation}

\begin{figure}[h]
\includegraphics[width=0.7\linewidth]{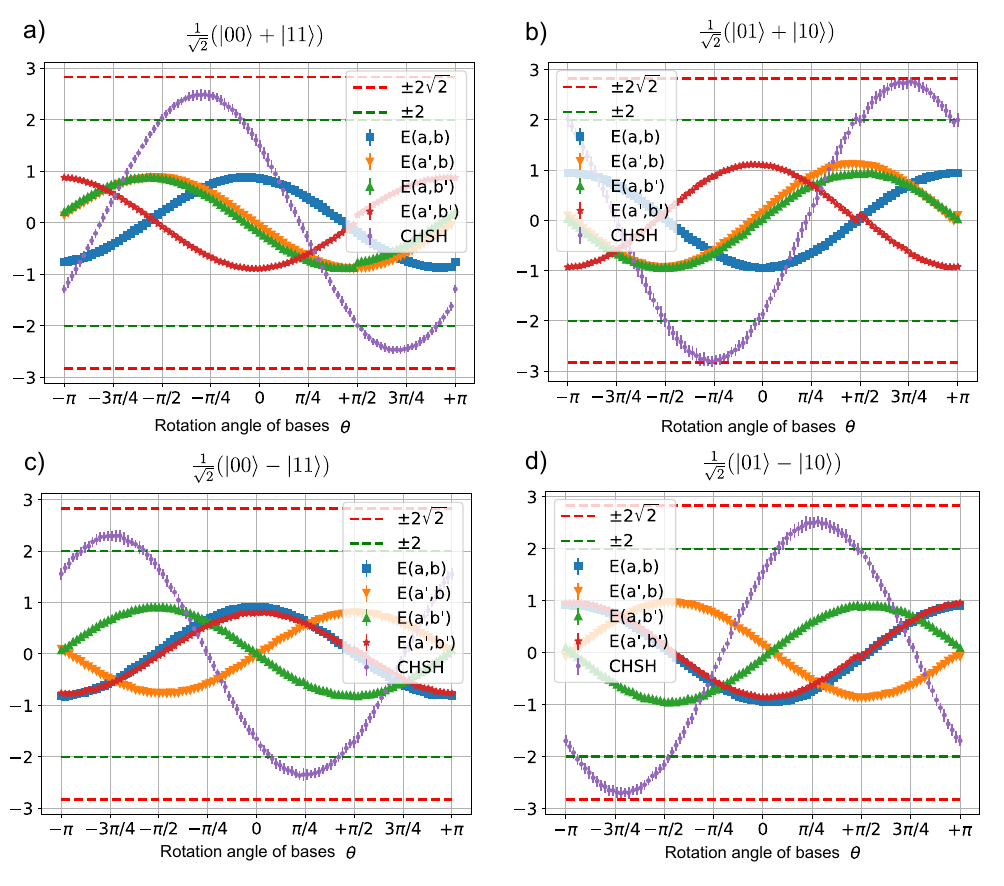}% Here is how to import EPS art
\caption{\label{fig:CHSH} Measured correlators and CHSH expression for 4 Bell states.}
\end{figure}

The Figure \ref{fig:CHSH} shows the measured violations of the CHSH inequality for all 4 prepared Bell states.

\clearpage

\bibliography{bibliography}